# Application of nonlinear noise regression in the Virgo detector

R. Weizmann Kiendrebeogo,[1,2,3,4,*] Muhammed Saleem,[5,3,†] Marie Anne Bizouard,[2] Andy H. Y. Chen,[6] Nelson Christensen,[2] Chia-Jui Chou,[7] Michael W. Coughlin,[3] Kamiel Janssens,[8,2,9,10] S. Zacharie Kam,[1] Jean Koulidiati,[1] and Shu-Wei Yeh[11]

[1]*Laboratoire de Physique et de Chimie de l'Environnement, Université Joseph KI-ZERBO, Ouagadougou, Burkina Faso*
[2]*Université Côte d'Azur, Observatoire de la Côte d'Azur, CNRS, Laboratoire Artemis, 06300 Nice, France*
[3]*School of Physics and Astronomy, University of Minnesota, Minneapolis, Minnesota 55455, USA*
[4]*IRFU, CEA, Université Paris-Saclay, F-91191 Gif-sur-Yvette, France*
[5]*Department of Physics, University of Texas at Austin, Austin, Texas 78712, USA*
[6]*Institute of Physics, National Yang-Ming Chiao Tung University, Hsinchu, Taiwan*
[7]*Department of Electrophysics, National Yang Ming Chiao Tung University, Hsinchu, Taiwan*
[8]*Universiteit Antwerpen, Prinsstraat 13, 2000 Antwerpen, Belgium*
[9]*Department of Physics, The University of Adelaide, Adelaide, South Australia 5005, Australia*
[10]*ARC Centre of Excellence for Dark Matter Particle Physics, Melbourne, Australia*
[11]*Department of Physics, National Tsing Hua University, Hsinchu, Taiwan*

(Received 16 April 2025; accepted 26 September 2025; published 23 October 2025)

This work presents the first demonstration of nonlinear noise regression in the Virgo detector using deep learning techniques. We use `DeepClean`, a convolutional autoencoder previously shown to be effective in denoising LIGO data, as our tool for modeling and subtracting environmental and technical noise in Virgo. The method uses auxiliary witness channels to learn correlated noise features and remove them from the strain data. For this study, we apply `DeepClean` to Virgo O3b data, using 225 witness channels selected across 13 targeted frequency bands. Our analysis confirms the presence of nonlinear couplings in the subtracted noise, highlighting the importance of DeepClean-like tools in capturing such effects. We observe up to a 1.3 Mpc improvement in the binary neutron star inspiral range (~2.5% gain), and an average increase of 1.7% in the recovered signal-to-noise ratio for injected binary black hole signals. Parameter estimation studies further confirm that `DeepClean` does not introduce bias in the recovery of source parameters. These results demonstrate the robustness of `DeepClean` on Virgo data and support its adoption in real-time noise subtraction frameworks for future observing runs.

## I. INTRODUCTION

Gravitational-wave (GW) events, such as GW150914 [1], GW170817 [2], GW190425 [3], and GW200105 [3], among others, have significantly improved our understanding of compact binary mergers [4,5]. However, many potential GW signals remain undetected just below the noise floor, awaiting the successful mitigation of environmental noise sources.

In the absence of a GW, the detectors may show fluctuations due to various factors that contribute to the noise of the system, including fundamental, technical, and environmental sources. These noise sources contribute with different and time-varying characteristics in amplitude and frequency. The characterization of the various noise components is crucial to understanding and improving the performance of a detector, in particular when identifying authentic GW signals [6,7]. Fundamental noise includes inherent limitations of detector design materials and quantum mechanics. Thermal noise in the mirror coatings is a fundamental noise in the frequency band of approximately 60–300 Hz. Environmental noise covers external influences like *seismic noise*, *atmospheric noise*, *external thermal fluctuations*, and *Newtonian noise* [8]. Technical noise comes from the design and operational aspects of the detector.

GW detectors are affected by various noise sources. Although the $h(t)$ readouts are similar, they differ in arm length (4 km for LIGO and 3 km for Virgo), electronic components, optical configuration, control design, and geographic location, which expose them to different operating conditions and to distinct noise environments that influence their sensitivities. In this work, we present the first application of DeepClean to Virgo data, a dedicated noise subtraction effort aimed at improving the detector sensitivity.

*Contact author: weizmann.kiendrebeogo@oca.eu
†Contact author: muhammed.cholayil@austin.utexas.edu

The presence of various noise sources make it difficult to clearly discern GW signals, mitigating or subtracting noise within the detectors may expose these missed signals to detection algorithms. To potentially reveal new signals or otherwise improve the SNR of known ones, both the Laser Interferometer GW Observatory (Advanced LIGO; [9]) and Advanced Virgo [10] record several thousands of witness channels (probes and sensors) [6,11]. These witness channels independently are sensitive to noise from various sources. Some of them collect data to characterize environmental noise sources that couple with the GW readout channel and could therefore be used to subtract noise sources.

Only nonfundamental noise can be subtracted. Thus, we prioritize mitigating measurable environmental and technical noise. Through the classification of noise into removable and nonremovable categories, our objective is to enhance the sensitivity of GW detectors, ensuring a clearer distinction between real GW signals and noise artifacts.

Noise subtraction methods such as Wiener filtering have been widely used to subtract linearly coupled noise sources, however they face limitations in the presence of *nonlinear* and *nonstationary* noise [12].

At present, there are several narrow-band and broadband frequency ranges where the sensitivity is significantly made worse than that expected by the design. Low-frequency (typically around 10–100 Hz) sensitivity is particularly important for many reasons [13,14]. These include the possible detections of high-mass binary mergers at low frequencies, stochastic GW background detection which significantly benefits from the excess GW power in low frequencies [15,16], and the premerger detectability of binary neutron star (BNS) mergers [2] as it allows us to anticipate potential electromagnetic (EM) counterparts [17,18]. Furthermore, many GW signals that are hidden in noise, would become detectable when noise is adequately minimized in any frequency band.

Much noise is caused by the presence of various environmental and technical noise contributions. It is possible that their sources might have already been being tracked by one or more of the witness sensors. Identifying the appropriate witness channels is a challenging task, which is often done by the experimentalist's intuition. There are also algorithmic approaches being developed for this purpose, which will be subject of a future paper.

DeepClean is a deep-learning infrastructure developed to denoise the GWs strain data using witness channels [13,19]. It has a 1D Convolutional Neural Network (CNN) architecture, with the time-series data from multiple witness sensors as inputs and a single noise prediction as the output. Due to its complex architecture, DeepClean is capable of modeling and subtracting *nonlinear* noise couplings from the GW data. DeepClean is also a lightweight architecture which takes only a few minutes to train on roughly one hour of data and few seconds to minutes to perform cleaning on several hours of data (for example, on a standard NVIDIA A100 Tensor Core GPU), potentially making the analysis time much smaller than the data duration. This makes it feasible to retrain DeepClean so often that the *nonstationary* features can be efficiently captured. The performance of DeepClean on LIGO detectors has already been demonstrated in detail in the aforementioned publications [13,19].

In this work, we focus on the Virgo detector, where some EM noise comes from electrical sources, reaching their peak at the 50 Hz power line frequency. This electrical noise induces modulations in the main GW signal, leading to the generation of symmetric sidebands around 50 Hz. During O3, a feed-forward loop process was used to remove the Virgo mains and its associated sidebands [20]. We also aim to demonstrate the flexibility of the architecture in to be applicable on detectors with different designs. In addition, in this work, we also demonstrate a multilayer multiband training approach (simply known as *multitraining*) unlike the subtractions done in [19] which were done for a single frequency band with a single layer of training.

In Sec. II, we describe the methodology implemented in the DeepClean algorithm, focusing on noise estimation and reduction using witness channels. Section III gives a discussion of the application of DeepClean on Virgo's O3b (the second part of the third observing run of LIGO-Virgo Collaboration) data, including the training process and noise subtraction. The results, including analyses of individual frequency bands and the multitraining approach, are presented in Sec. III C. Our conclusion and future prospects are summarized in Sec. V.

## II. DEEPCLEAN AS A TOOL FOR NONLINEAR NOISE REGRESSION

Several noise regression techniques have been developed to mitigate environmental and instrumental disturbances in GW detectors. These include classical linear methods such as Wiener filtering [12,21,22], and adaptive feed-forward control schemes [20], as well as more recent machine learning (ML)-supported parametric approaches like NonSENS [14]. Beyond these, a growing body of work explores ML-based denoising strategies [23–27], positioning DeepClean within the evolving landscape of data-driven noise regression. Moreover, ML is also increasingly applied to the identification of electromagnetic counterparts of GW sources [28–30].

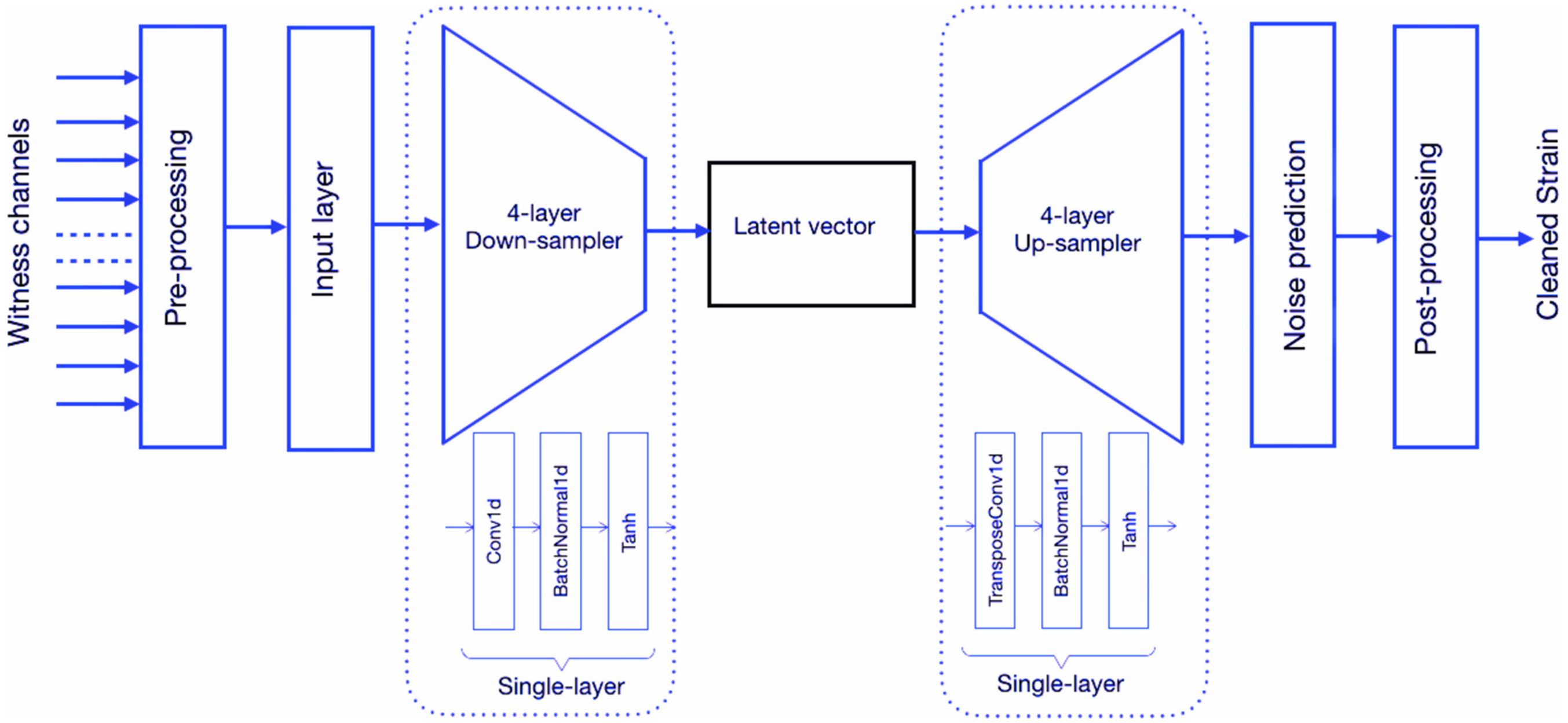

FIG. 1. Architecture of the DeepClean system and the associated data processing sequence. DeepClean processes input in the form of time series data acquired from an array of witness sensors. The data undergo a transformation through a convolutional autoencoder network, which consists of a series of four convolutional layers for the reduction of dimensionality, followed by an equal number of transpose-convolutional layers responsible for dimensionality expansion. Each convolutional operation is succeeded by batch normalization and hyperbolic tangent (tanh) activation. The final convolutional layer culminates in a one-dimensional representation aimed at predicting noise.

See Appendix A 1 for a short summary of some of the popular methods. While each method has its advantages, recent developments have shown that deep learning tools offer a powerful framework for capturing complex, nonlinear, and potentially nonstationary noise couplings. In this study, we use DeepClean, a convolutional autoencoder previously demonstrated in LIGO data [13,19], to perform nonlinear noise regression in the Virgo detector. We describe the architecture and training methodology of DeepClean in detail below.

DeepClean uses a CNN architecture to measure the correlation between different auxiliary sensors that is coupled to the GW data [13,19], to mitigate environmental and instrumental noise off the GW data. Figure 1 summarizes the workflow. The architecture features a symmetric one-dimensional autoencoder with multiple downsampling and upsampling layers, tailored to the sampling frequency and witness channel count. This configuration enables noise prediction across different data dimensions and uses batch normalization and tanh activation functions to boost generalization. The trainable weights of the neural network determine the form of the noise coupling. In the most general case, it can capture *nonlinear* couplings by virtue of the nonlinear activation functions included in the neural network [13]. With the help of auxiliary sensors tracking the changes in coupling features over time, it can in principle also capture *nonstationary* noise couplings. The training minimizes a loss function based on the noise spectrum ratio of cleaned to original strain across all frequency bins.

The GW strain $h(t)$ from a detector can be written as follows:

$$h(t) = s(t) + n(t), \tag{2.1}$$

where $s(t)$ denotes potential astrophysical signals and $n(t)$ is the detector noise. The noise can be further decomposed as

$$n(t) = n_R(t) + n_{NR}(t), \tag{2.2}$$

where $n_{NR}(t)$ is known as the nonremovable noise which are the inherent limitations of the design.[1] On the other hand, $n_R(t)$ is called the removable noise which is not there in the core design of the interferometers but originates from environmental and technical factors.

The interferometers are equipped with thousands of auxiliary witness channels that record independent instances of noise from various technical and environmental sources and are used to reconstruct the removable noise $n_R(t)$. Based on the physical origin, the $n_R(t)$ can be understood to be a superposition of several noise couplings, schematically written as $n_R(t) = n_R^{(1)}(t) + n_R^{(2)}(t) + ..$, with each term being the result of a unique noise coupling between a unique subset of the witness channels. The denoising efforts with DeepClean discussed in this work have to be applied separately on each of those contributions. Our efforts here are to mitigate one or more of the contributors rather than subtracting the entire $n_R(t)$, which is beyond the scope of this work.[2] Henceforth in this paper, $n_R(t)$ indicates those couplings which we intend to subtract in this work.

[1]For example, the dependency of thermal noise on the mirror coating material, and the photon shot noise on the laser power used.

A convenient mathematical notation for the DeepClean neural network is $\mathcal{F}(w_k;\vec{\theta})$, where $w_k(t)$ represents the data from the witness sensors, and $\vec{\theta}$ is the set of all the trainable weights of DeepClean, which are optimized by minimizing a loss function $J_{\mathrm{PSD}}$,

$$\begin{cases} n_R(t) = \mathcal{F}(w_k, \vec{\theta}), \\ \vec{\theta} = \mathrm{argmin}_{\theta'} J_{\mathrm{PSD}}[h(t), \mathcal{F}(w_i(t); \vec{\theta'})]. \end{cases} \tag{2.3}$$

Subtracting $n_R(t,\vec{\theta})$ from the GW strain data $h(t)$ yields the residual strain $r(t,\vec{\theta})$ also known as cleaned strain,

$$r(t,\vec{\theta}) = h(t) - n_R(t,\vec{\theta}). \tag{2.4}$$

As mentioned before, the cleaned strain would still contain the nonremovable noise, the unknown contributions to the removable noise, and any possible astrophysical signals that are present. The loss function $J_{\mathrm{PSD}}$ is defined as the ratio of the power spectral densities of the cleaned and raw strain summed over all the frequency bins within the subtraction band,

$$J_{\mathrm{PSD}} = \frac{1}{N}\sum_{k=1}^{N}\frac{\mathrm{PSD}_r[k]}{\mathrm{PSD}_h[k]}, \tag{2.5}$$

where $\mathrm{PSD}_h[k]$ represents the power spectral density (PSD) derived from the original strain data, and $\mathrm{PSD}_r[k]$ represents the PSD of the residual strain, $r(t)$. The index $k$ runs over all the $N$ frequency bins, uniformly spaced between $f_{\mathrm{low}}$ and $f_{\mathrm{high}}$. To optimize the weights ($\vec{\theta}$), DeepClean employs the ADAM optimizer [31].

## III. A VIRGO O3B MOCK DATA CHALLENGE

We apply DeepClean on Virgo O3b data and assess the performance using various metrics. We examined a continuous stretch of GW strain data `Hrec_hoft_raw_20000Hz` and the useful witness channels, spanning two days, 18 h and 15 min, from February 7, 2020, 16:19:27 UTC to February 10, 2020, 10:35:01 UTC. This period was selected for its extensive coverage and the significant coherence observed between strain and auxiliary channels in the Virgo O3b dataset which is essential for the effective deployment and evaluation of DeepClean.

The models were trained on segments of 1024 s, 2048 s, and 4096 s durations, with the segment length empirically chosen to balance computational cost with training stability. In practice, longer training segments tend to yield more robust learning by emphasizing persistent features of the noise couplings rather than short-term variations. To account for evolving coupling behavior and auxiliary channel reliability, the models were retrained approximately once every $10^5$ s of data. Beyond this cadence, subtraction performance degraded noticeably, necessitating periodic retraining.

Each model was used to clean the corresponding segment length. We first focus noise reduction specifically within three frequency bands; 98–110 Hz, 142–162 Hz, and 197–208 Hz. We target these bands due to their excessive noise, which is appropriate for testing the effectiveness of the algorithm and its ability in enhancing the recovery of astrophysical signals in the data. Noise reduction in these bands are particularly interesting in the context of premerger detections of compact binary coalescence (CBC) events, especially binary neutron star mergers and other low-mass black hole binaries.

### A. Witness sensors

A comprehensive analysis of witness channels was first performed within the specific frequency bands. To determine which channels are the most important, we consider the coherence between witness channels and the GW strain channel. We consider the output of the so-called BruCo algorithm (a brute-force coherence computation tool among all channels in Virgo; [32]), which run automatically during the O3b, to select channels whose coherence exceeded 0.5 (50% of the maximum possible coherence). Additionally, we required that this coherence excess occurred at least five times during February and March 2020. This ensures that each witness channel shows sustained coherence levels across multiple instances. Each witness channel is tested for its noise removal effects and those deemed to not contribute are removed. The choice of a 0.5 threshold represents a tradeoff; lowering the threshold would include many weakly correlated channels, which in our tests did not improve the subtraction performance but significantly increased the training complexity and computational cost. Conversely, setting the threshold too high risks excluding moderately coupled channels that still provide useful noise regression. The adopted criterion therefore balances computational efficiency with denoising performance, ensuring a manageable set of channels with consistent contribution. Below we describe the main witness channels selected for each frequency band.

[2]Note that there are several frequency bands with unknown noise, which are neither due to design limitations nor associated with known witness channels. Although such noise is included in $n_R(t)$ and falls under the category of removable noise, it is not currently removable given our present understanding of the available witness channels.

98 to 110 Hz:

(i) `V1:CAL_WE_MIR_Z_NOISE`: Injects and monitors longitudinal noise in the west end mirror for calibration.

(ii) `V1:CAL_NE_MIR_Z_NOISE`: Injects and monitors longitudinal noise in the north end mirror for calibration.

(iii) `V1:INJ_IMC_QD_FF_DC_V`: Measures the DC voltage of the Input Mode Cleaner.

197 to 208 Hz:

(i) `V1:CAL_WE_MIR_Z_NOISE`: Injects and monitors longitudinal noise in the west end mirror for calibration.

(ii) `V1:SDB_EDB_Tpro_processed_packets` and `V1:SDB2_Tpro_processed_packets`: Impacts data quality via suspension database processing.

(iii) `V1:ENV_IB_CT_FINGER_ACC_Y`: Monitors environmental vibrations.

(iv) `V1:ENV_CEB_MAG_W`: Monitors magnetic fields and detects magnetic noise.

(v) `V1:INJ_IMC_QD_FF_I_H` and `V1:INJ_IMC_REFL_I_POST`: Injection Mode Cleaner control signal from the in-phase demodulated quadrant photodiode.

(vi) `V1:SQZ_CC_Tpro_processed_packets`: Squeezing control system packets.

142 to 162 Hz:

Given the complex contributions of 167 witness channels in this frequency range to noise reduction, a full listing is omitted. Key acronyms like ASC (alignment sensing and control), INJ (injection), ACT (actuation), and ENV (environment) are emphasized to highlight the main systems that maintain the operating integrity of the interferometer [6]. The complete list of witness channels can be found here.[3]

### B. Training and cleaning

For each frequency band described above, training is performed independently. What this means is that three separate instances of DeepClean are trained, one for each band with the respective set of channels. This leads to three different versions of cleaned strain data with each of them achieving noise reductions in the respective bands. There is no version of cleaned strain with subtraction achieved in all three bands. In a more practical scenario, for a production analysis, training on each band should be performed sequentially, with the cleaned strain output from one band being considered as the unclean input strain for the next band, leading to the final layer giving cleaned strain with noise subtraction achieved in all the considered bands. This has been carried out and described below, with not only three but many more frequency bands and witness channels. For the moment, we focus on the performance in each band separately.

[3] https://github.com/weizmannk/Virgo-DeepClean/blob/main/config/witnesses/single-layers/witnesses_142-162_Hz.ini

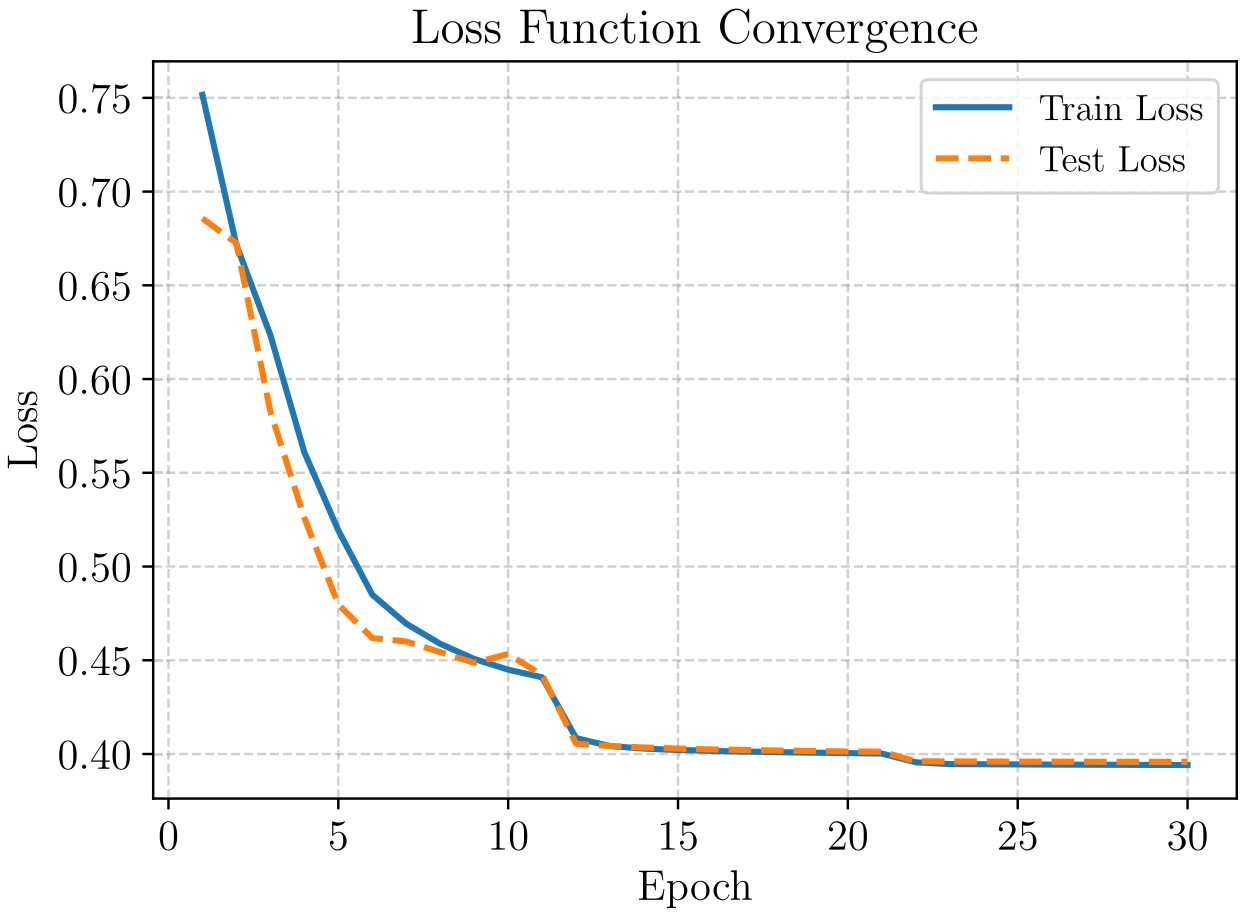

FIG. 2. The convergence of the loss function during training is shown. The solid blue line represents the training loss, while the dashed orange line indicates the test loss over 30 epochs. Both curves show consistent convergence, demonstrating effective learning and minimal overfitting.

There are a number of hyperparameters which can optimize the performance of DeepClean, including the learning rate, kernel size, stride etc. For all of these, we use the same settings as used in [19]. The data from all the witness channels are first preprocessed such that each time series has zero mean and unit variance. These normalized datasets are bandpassed and then divided into overlapping kernels.[4] with kernel size of 8 s, and 7.75 s overlaps. The 4096 s of training data yields 16353 overlapping kernels, calculated using the formula,

$$\text{kernels} = \frac{\text{data duration-kernel duration}}{\text{shift}} + 1 = \frac{4096 - 8}{0.25} + 1.$$

These kernels are then grouped into batches of 32 (known as the `batch_size`), resulting in a total of 512 batches. This implies that during one epoch of training, DeepClean iterates and updates the weights 512 times. When the witness channels exhibit moderate to strong coupling, the weights typically converge to their optimal values within 20–30 training epochs. Conversely, if the coupling is absent or very weak, the training may fail to converge or require additional hyperparameter tuning to achieve meaningful results. Throughout this study, we have used a fixed number of 30 epochs for training. Figure 2 shows the loss in relation to the epochs.

[4] The kernels are here the segments of data and are distinct from the convolution filter kernels in CNN architecture that carry the weights.

TABLE I. Summary of the DeepClean training configuration with Virgo O3b data. The full configuration files are publicly available at[a].

| Parameter | Value |
|---|---|
| *Data segmentation* | |
| Sampling frequency | 4096 Hz |
| Length of overlapping kernels | 8 s |
| Stride (training / cleaning) | 0.25 s/4 s |
| *Preprocessing* | |
| Bandpass filter (8th order, Hanning window, median padding) | |
| *Training settings* | |
| Optimizer | Adam |
| Batch size | 32 |
| Epochs | 30 |
| Learning rate | $10^{-3}$ |
| Weight decay | $10^{-5}$ |
| Training duration | 4096 s |
| Training cadence | $10^5$ s |
| At-a-time cleaning duration | 4096 s |
| *Dataset* | |
| Detector run | Virgo O3b |
| Training/validation data split | 90% (10% validation) |

[a]https://github.com/weizmannk/Virgo-DeepClean/blob/main/config/config_virgo.ini

The cleaned strain is produced in chunks of 4096 s for all three bands. In addition to the normalization of the input witness channels mentioned earlier, there are batch normalizations applied at each layer of the neural network to ensure consistent input distributions for the subsequent layers. In DeepClean, batch normalization also makes sure that the final noise predictions have nearly zero mean and unit variance. For this reason the predicted noise is subtracted from normalized version of the raw strain, which yields cleaned strain in normalized units. This is then unnormalized with the mean and standard deviation used earlier to normalize the unclean strain. Following that, the cleaned strain is bandpassed to the target frequency band, in order to make sure that DeepClean did not make any noise prediction outside the desired frequency band. Hereafter in this paper, *cleaned strain* refers to this postprocessed cleaned strain. Table I presents the configuration used for training and cleaning.

## C. Performance evaluation with amplitude spectral density

### 1. Single layer subtractions and their amplitude spectral density ratio

The primary approach used here for assessing DeepClean's performance is the comparison of the amplitude spectral density (ASD) of the cleaned strain (labeled V1:DC) against the original unclean strain (labeled V1:ORG), where V1 denotes the Virgo detector. Figure 3 illustrates the ASD enhancements for the targeted frequency ranges of 98–110 Hz, 142–162 Hz, and 197–208 Hz. The upper panels in each set of plots display the ASDs, while the lower panels show the ASD ratios of the cleaned strain to the uncleaned strain. Significant subtractions are observed across all three bands. The efficacy of DeepClean varied significantly with the length of the training data. Although segments of 1024 s and 2048 s were used, the results were suboptimal compared to those obtained from the 4096 s segments.

This may be attributed to instabilities present in some witness channels during O3b, where sensors intermittently failed to record data, resulting in periods of low or zero coherence with the strain channel. Shorter training segments are more likely to overfit to these unstable intervals, while longer segments promote learning of persistent coupling patterns. This is consistent with the improved subtraction performance observed for 4096 s training compared to shorter durations. Specifically, Fig. 3 shows that the model trained with 4096 s segments achieves more than 99.9% suppression at the dominant narrow peaks (top panel), more than 88% suppression across all peaks (bottom panel), and a median reduction of about 72% in the 149–152 Hz band (middle panel).

### 2. Multitraining and ASD ratio

This section introduces a segmented multitraining approach targeting specific frequency bands that collectively cover the range from 15 to 415 Hz, with each band trained independently.

The multitraining methodology applies a sequential, layerwise approach. The training and noise cleaning are performed iteratively across frequency bands, with the output of one layer serving as the input for the next, progressively refining noise reduction across the different frequency bands. The process begins with the 142–162 Hz band, selected for its 122 associated witness channels, requiring at least 80 GB of RAM to process 4096 s of Virgo O3b data. This initial layer's cleaned output forms the basis for subsequent layers.

The full process comprises 13 layers in total; 15–20 Hz, 33–39 Hz, 55–65 Hz, 75–80 Hz, 98–110 Hz, 137–139 Hz, 142–162 Hz, 197–208 Hz, 247–252 Hz, 295–305 Hz, 345–355 Hz, 355–367 Hz, and 395–415 Hz band. Each frequency band is associated with specific witness channels[5] that demonstrate a certain level of coherence. The ASD for the original and cleaned data is presented in Fig. 4.

[5]https://github.com/weizmannk/Virgo-DeepClean/tree/main/config/witnesses/multi-layers

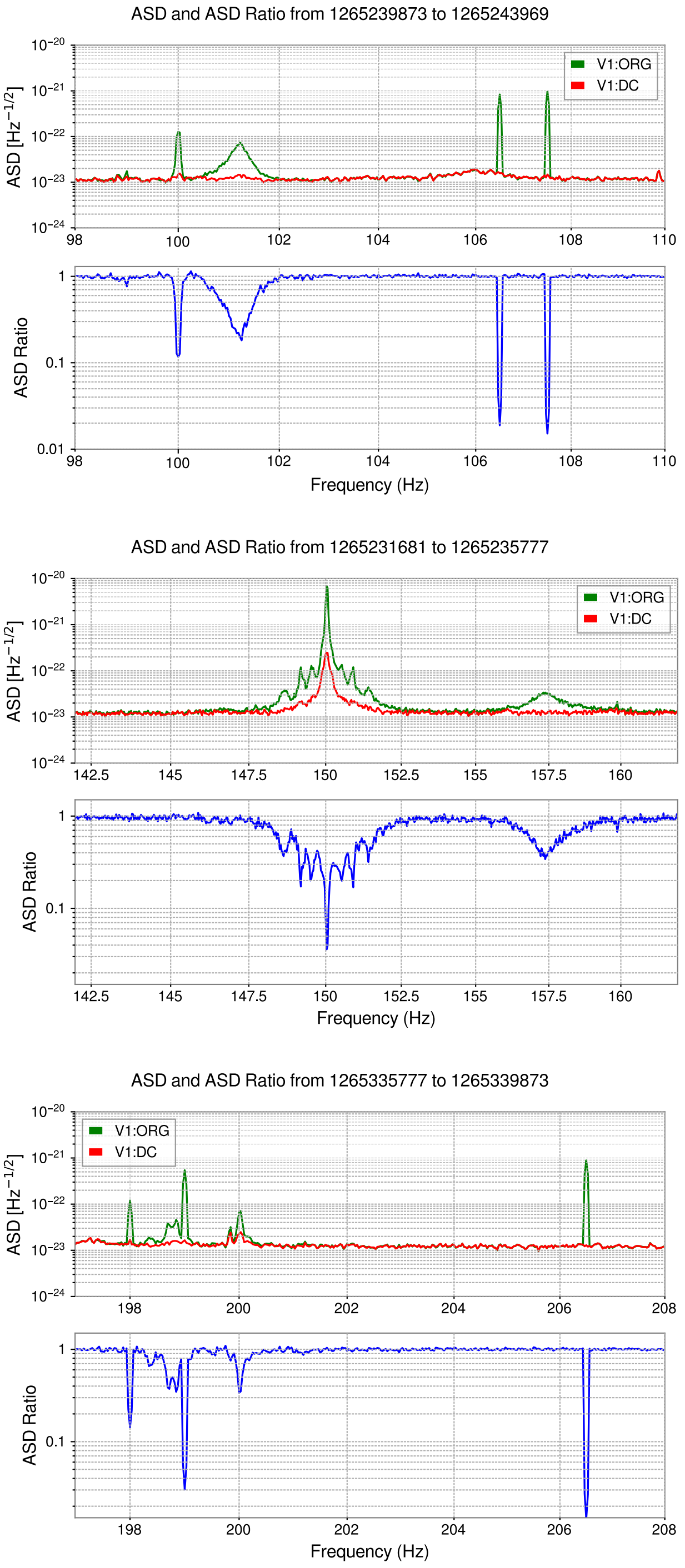

FIG. 3. ASD and ASD ratio comparison across different frequency bands for Virgo detector data. The green curve represents the ASD of the original data from Virgo (V1:ORG), the red curve depicts the ASD after processing with the DeepClean algorithm (V1:DC), and the blue curve shows the ratio of the two ASDs. The top panel demonstrates noise reduction in the 98–110 Hz frequency band, the middle panel in the 142–162 Hz band, and the bottom panel in the 197–208 Hz band. These panels demonstrate the noise reduction capabilities of DeepClean within this specific frequency range.

### D. Performance evaluation with binary neutron star inspiral range

The application of the DeepClean algorithm for noise subtraction in GW detectors yields quantifiable improvements in the binary neutron star (BNS) inspiral range, which is a measure of the average distance at which a BNS system with component masses of $m_1 = m_2 = 1.4M_\odot$ can be detected with a SNR of 8. This range effectively serves as an indicator of overall GW detector sensitivity [33–35]. An extended BNS inspiral range reflects heightened detector sensitivity, a direct result of advanced noise reduction methods that enable the capture of fainter, more distant signals.

#### *1. BNS range from single layer*

A comparative analysis conducted before and after implementing DeepClean reveals modest yet statistically significant enhancements. Specifically, within the 98–110 Hz band, an average increase of approximately 0.2 Mpc observed, corresponding to an improvement of 0.4%. In the 142–162 Hz band, the average gain approximately 0.19 Mpc or 0.4%. Furthermore, cleaning within the 197–208 Hz band results in a marginal increase of approximately 0.02 Mpc (0.05% improvement). Figure 5 illustrates the evolution of BNS inspiral range before and after noise subtraction, particularly within the 142–162 Hz band.

This BNS inspiral range shows a daily fluctuation, with maximal values occurring around midnight and at minimal values after 9:00 AM (UTC), driven by local environmental noise sources and site-specific dynamics. These fluctuations primarily stem from changes in anthropogenic noise (e.g., traffic vibrations) and wind conditions [36,37]. Nighttime brings optimal sensitivity due to minimal human activity and lower seismic noise, while morning traffic increases ground vibrations, reducing sensitivity.

We will revisit this analysis in a later section, incorporating results from subtraction performed across multiple bands.

#### *2. BNS range from multitraining*

This approach has achieved a significant improvement in the BNS inspiral range, increasing it by approximately 1.3 Mpc, representing a 2.5% enhancement. Similar to the single training case, the contribution of different frequency bands to the inspiral range varies significantly. Some bands contribute substantially, while others have little to no impact. In our analysis, the 55–56 Hz band contributes the most, with an increase of 0.51 Mpc, followed by the 33–39 Hz band, which adds 0.29 Mpc. Frequency bands such as 75–80 Hz, 98–110 Hz, and 142–162 Hz contribute modestly, with increases ranging from 0.12 Mpc to 0.17 Mpc. Other bands show minimal or negligible contributions. Figure 6 illustrates the BNS evolution as function of the frequency band (layer index). The results not only reveal the individual contribution of each frequency band but also highlight the importance of this sequential training process. Such a process could significantly enhance the sensitivity of the Virgo detector.

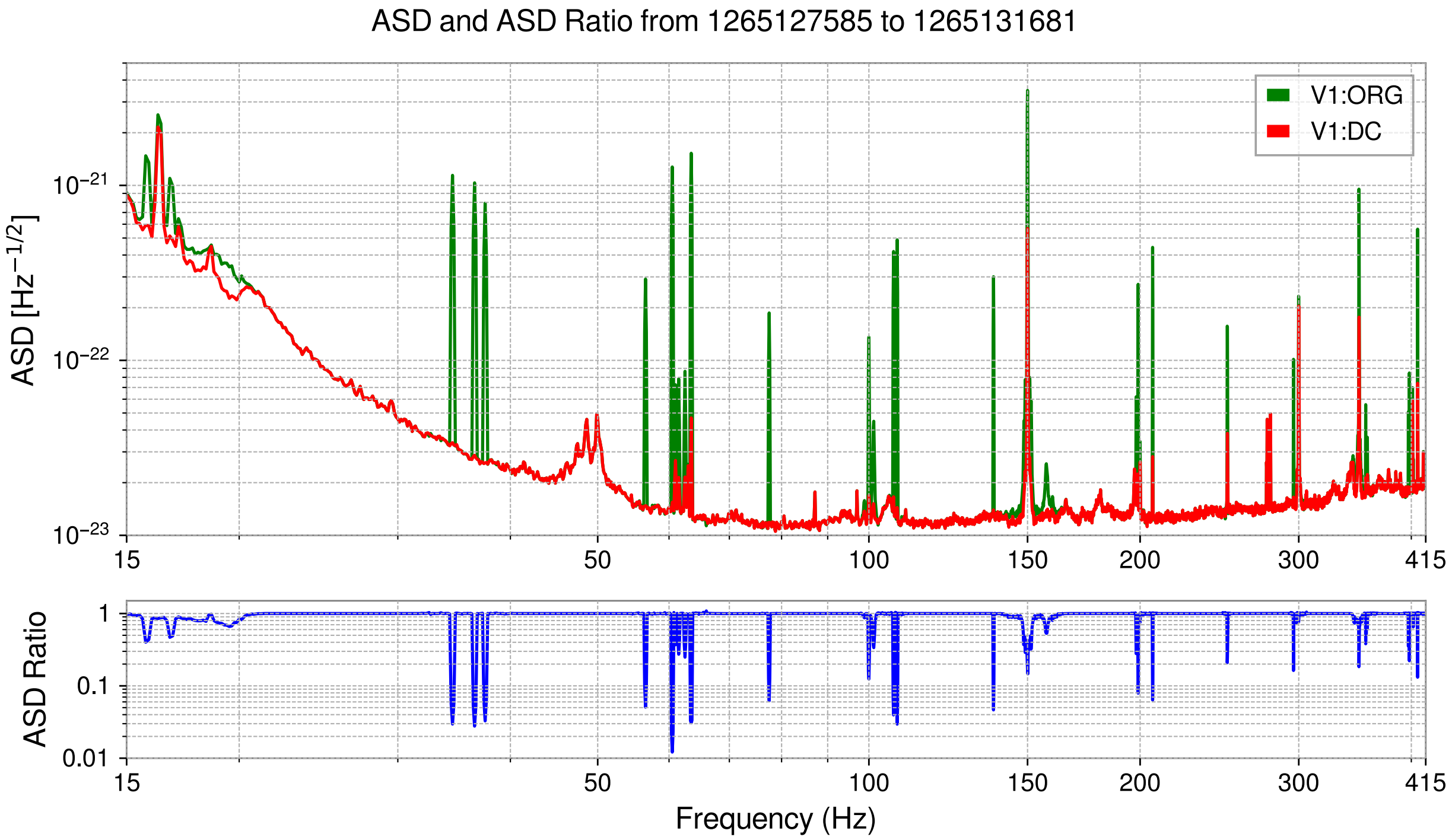

FIG. 4. Comparison of ASD and ASD ratio across a 15 to 415 Hz frequency band. The original data from Virgo is represented by the green curve (V1:ORG), whereas the red curve (V1:DC) displays the ASD after processing with the `DeepClean` algorithm. The blue curve indicates the ratio of the two ASDs. These visualizations highlight the `DeepClean` algorithm's noise reduction efficacy across the specified frequency spectrum.

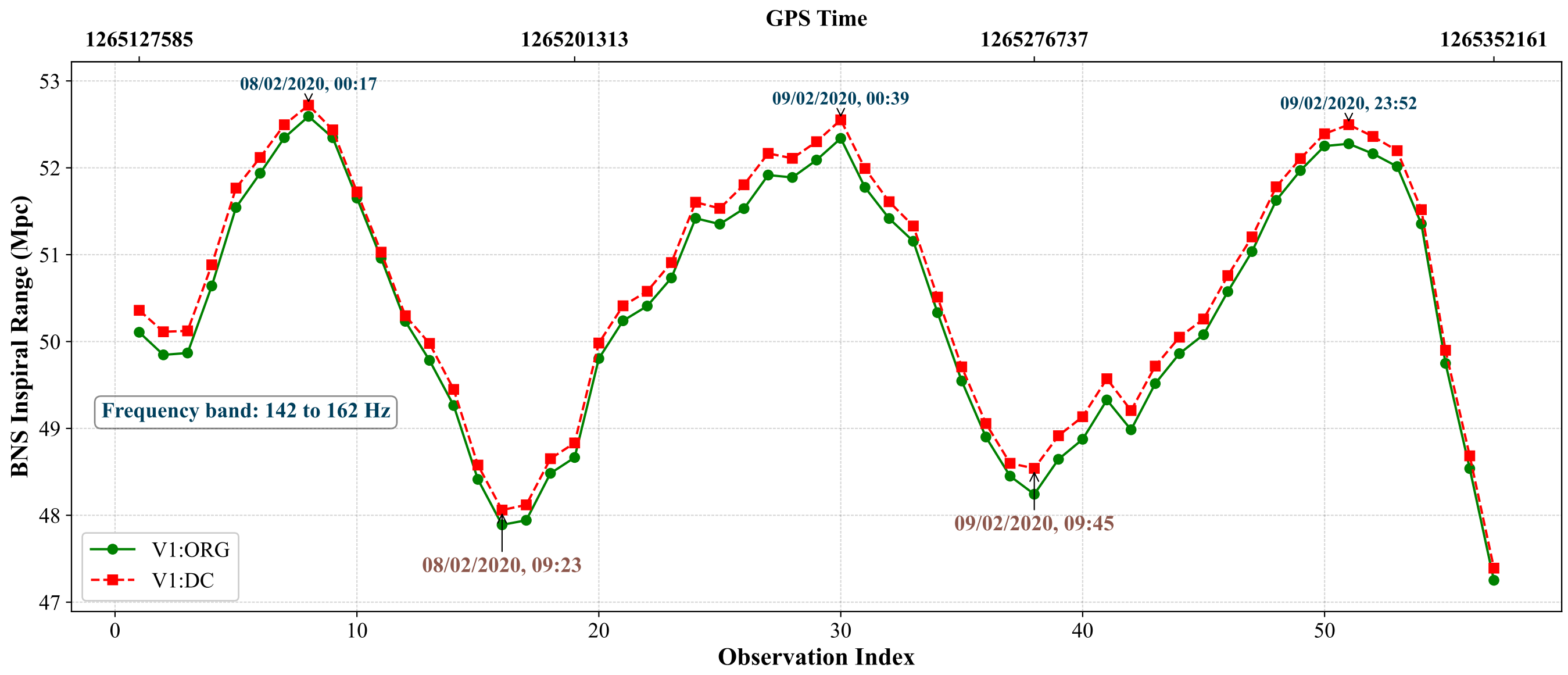

FIG. 5. Analysis of the inspiral range sensitivity for BNS, using Virgo O3b data from GPS time 1265127585 to 1265352161. The target frequency band for this analysis is the 142–162 Hz range. This plot highlights the `DeepClean` effectiveness in enhancing BNS inspiral detection over each 4096 s segment, represented by the "Observation Index." Note: The periodic fluctuations reflect anthropogenic noise near the Virgo detector, with the highest sensitivity around midnight and the lowest around 09:00 (UTC), when anthropogenic noise is most prevalent.

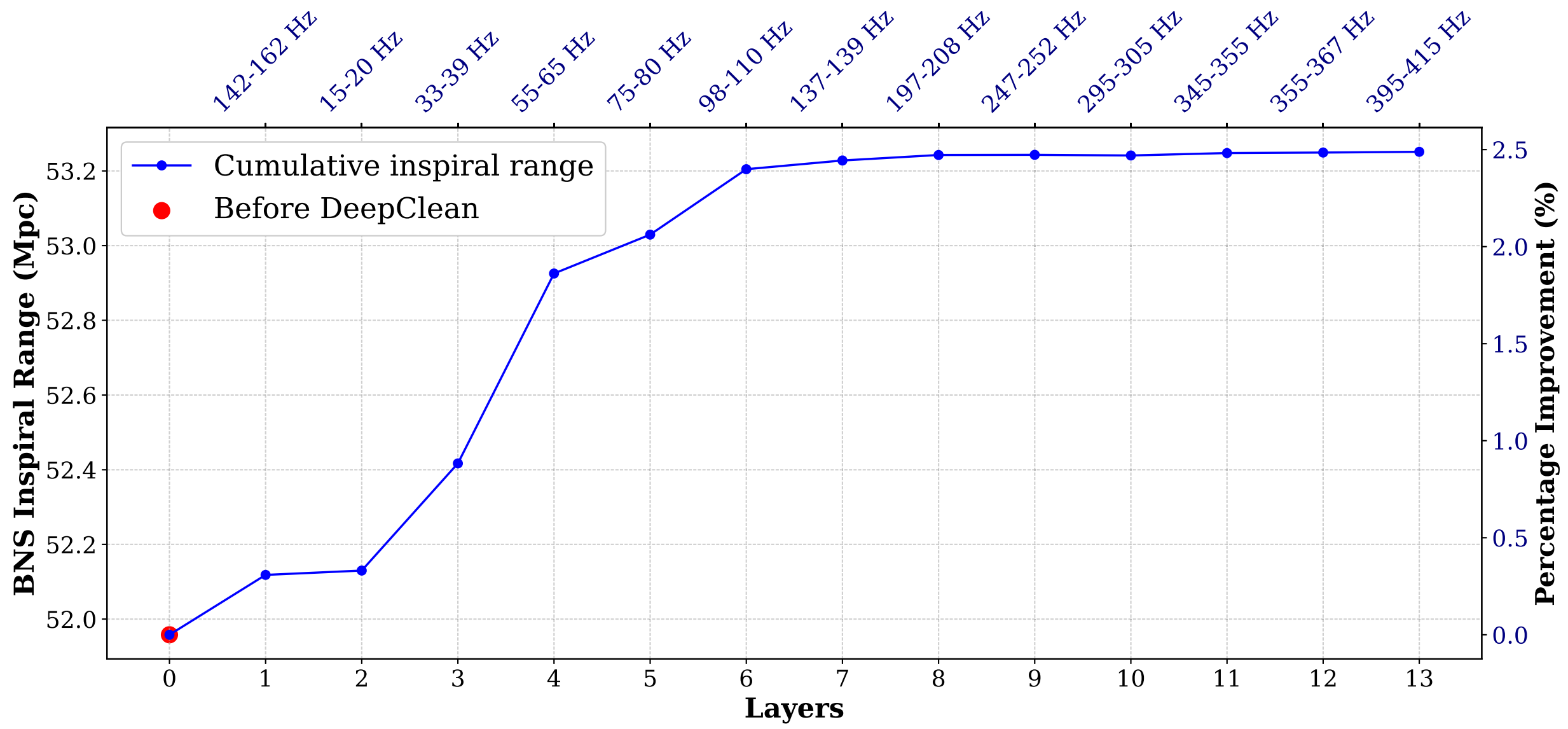

FIG. 6. Cumulative BNS inspiral range (blue curve) as a function of sequential training layers in the `DeepClean` algorithm, covering the frequency band from 15 to 415 Hz. Each layer corresponds to a distinct frequency band, starting with the 142–162 Hz band. Subsequent layers are trained iteratively, building upon the output of the preceding layer. The red marker represents the baseline inspiral range before applying `DeepClean`.

### E. On the nature of the subtracted noise: Linear vs nonlinear

To determine whether the Virgo noise subtracted by `DeepClean` includes nonlinear couplings, we designed a dedicated test targeting the role of nonlinear activation functions. Specifically, we trained two versions of the same network architecture using all 122 witness channels in the 150 Hz subtraction band; one with the standard `tanh` activation functions after each convolution and batch normalization layer (refer to Fig. 1), and another with these activations removed. While the resulting network without `tanh` is not strictly linear, it is significantly restricted in its ability to model nonlinear couplings.

We show the results in Fig. 7, left panel, where both models were able to reduce the central peak at 150 Hz, which is known to be a linearly coupled feature (with the model without activations performing slightly better). However, only the model with nonlinear activation functions effectively removed the adjacent sidebands, suggesting that their subtraction required modeling nonlinear couplings between witness channels. This result does not represent a comprehensive classification of the linear or nonlinear nature of the Virgo noise, but it provides strong evidence that at least some of the subtracted noise is nonlinear. More importantly, it demonstrates that with the existing set of Virgo witness channels, `DeepClean` is capable of learning and removing nonlinear noise couplings.

Beyond that, there are practical advantages to using `DeepClean`-like tools even in the case of linear noise. In a separate study, we tested two training strategies using the same 122 witness channels: (1) *single-pass* training, where all channels are used together in a single training session; (2) *multilayer* training, where the channels are divided into eight groups containing 14, 32, 32, 12, 8, 5, 9, and 10 channels, respectively. These groups were formed based on channel types—such as magnetic, electric, seismic, and alignment sensing and control—and processed sequentially. In the multigroup strategy, the cleaned strain from one group is passed forward to the next. The results, shown in Fig. 7, right panel, indicate that the *single-pass* strategy consistently outperforms the *multilayer* approach. This demonstrates that multilayer subtraction remains suboptimal, even when each group contains many channels.

The suboptimality arises because the optimization within each group is local and does not account for inter-group dependencies. That is, the coupling coefficients for one group are learned assuming fixed values (zero) for the coefficients associated with the remaining channels, thereby restricting the optimization to a local minimum in the loss function. In contrast, the *single-pass* strategy performs a joint optimization over the full parameter space, increasing the likelihood of reaching a global minimum.

This illustrates the ability of deep learning tools like `DeepClean` to perform optimal regression over high-dimensional input spaces, enabled by hardware-accelerated training on GPUs. Traditional analytical linear methods, such as Wiener filtering, become increasingly difficult to scale due to computational and numerical limitations when many channels are involved. Deep learning frameworks allow all relevant channels to be used simultaneously, enhancing performance regardless of whether the underlying couplings are linear or nonlinear.

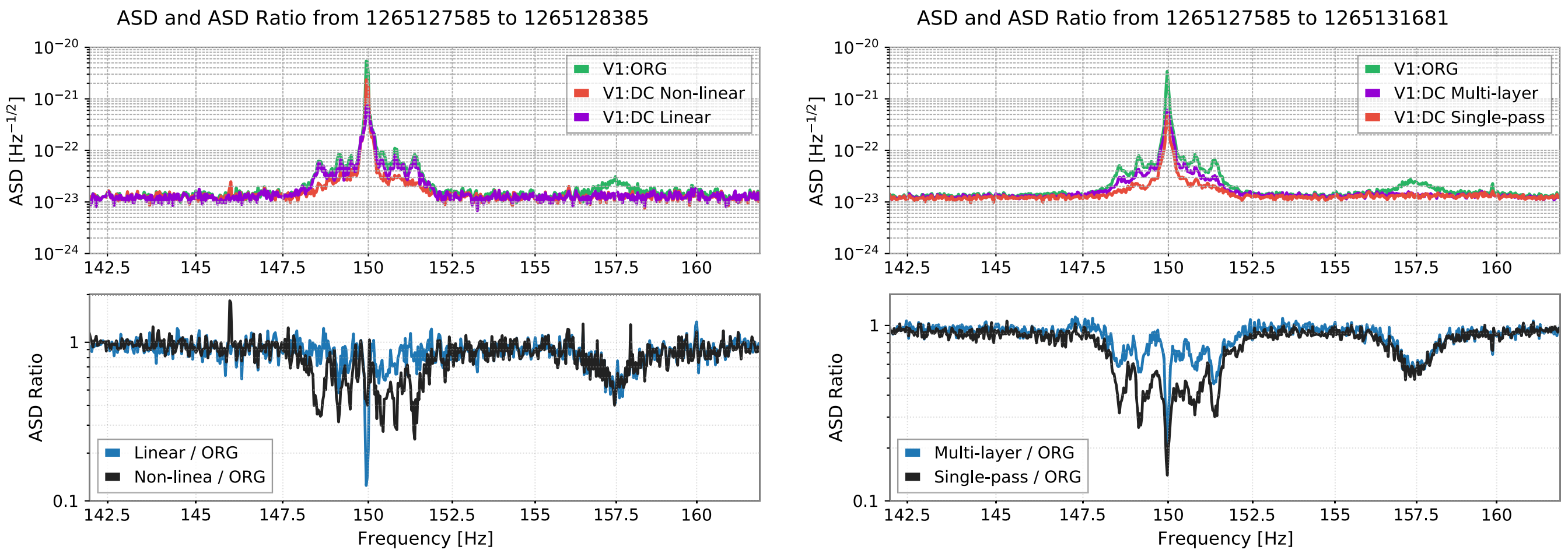

FIG. 7. Left: Comparison of the ASD and ASD ratios between the original Virgo data (V1:ORG) and the data processed with two versions of the `DeepClean` network; one with standard nonlinear `tanh` activations (red), and one with activations turned off (purple). Both models reduce the 150 Hz line, a linearly coupled feature, but only the nonlinear model reduces the nearby sidebands—suggesting that their subtraction requires modeling nonlinear noise couplings. Right: Comparison between two training strategies for `DeepClean` using all 122 witness channels: *single-pass* (red), which jointly optimizes over all channels in a single training session, and *multilayer* (purple), where the channels are divided into eight groups and processed sequentially. The *single-pass* model consistently outperforms the *multilayer* strategy, highlighting the importance of joint optimization across all channels to capture interchannel dependencies and avoid local minima. The bottom panel shows the corresponding ASD ratios relative to the original data.

## IV. VALIDATION ANALYSIS USING CBC INJECTIONS

We now conduct an injection analysis using CBC signals. The purpose is to verify that the denoising with `DeepClean` does not alter the GW signals that are present in the data, and also to examine any improvements in the credible intervals of the estimated parameters [38].

### A. GW generation and event selection

In the following, we use utilities LALSuite and Bilby libraries [39–41] to generate binary black hole (BBH) signals. The component masses $m_1$ and $m_2$ are drawn from a uniform distribution, ensuring that the total mass $M$ corresponds to innermost stable circular orbit (ISCO) frequencies $f_{\rm ISCO}$ falling within the three targeted frequency bands described above. A similar criterion was employed also by [19] for distributing the injection parameters.

The $f_{\rm ISCO}$ frequency is defined as

$$f_{\rm ISCO} = \frac{c^3}{6^{1.5}\pi GM}, \tag{4.1}$$

where $c$ is the speed of light, $G$ the gravitational constant, and $M$ the total mass of the binary system.

We inject BBH coalescence signals into the Virgo O3b raw data, referred to as our Mock Data. These injections had ISCO frequencies ranging from $f_{\rm low}$ to $f_{\rm high}$ and were spaced 32 s apart. The luminosity distances of the signals were uniformly distributed between 5 Mpc and 30 Mpc to ensure they were detectable. We apply `DeepClean` to both the original data and the data with injected GW signals.

### B. Signal-to-noise ratio analysis

Following the procedure described in Sec. IV A, we inject a total of 3,200 GW signals from BBH mergers into the dataset, spanning the full frequency range of 15–415 Hz. The SNR of each injected signal was computed using the matched-filter function provided by *PyCBC* [42], with a waveform template corresponding to the injected signal. For each injection, the matched-filter SNR was calculated both before and after applying `DeepClean`. In both cases, the filtering step used the same injected waveforms as templates, ensuring a direct comparison.

The difference in SNR is determined as the fractional change, $\frac{\Delta \rm SNR}{\rm SNR}$, where a positive value indicates a gain. Figure 8 displays the SNR fractional difference from multiband subtraction over the full range of 15–415 Hz, while Fig. 9 shows the distribution of fractional SNR differences for the 98–110 Hz and 142–162 Hz bands. We observe that some injections exhibit an SNR gain, while others show a loss. This variability is expected due to the stochastic nature of the noise prediction by `DeepClean`, which can influence the matched-filter SNR in either direction. However, the PSD consistently improves after applying `DeepClean`, contributing positively to SNR enhancement. The net effect on SNR arises from the combination of PSD improvement and the random nature of noise prediction, determining whether an injection experiences a gain or loss.

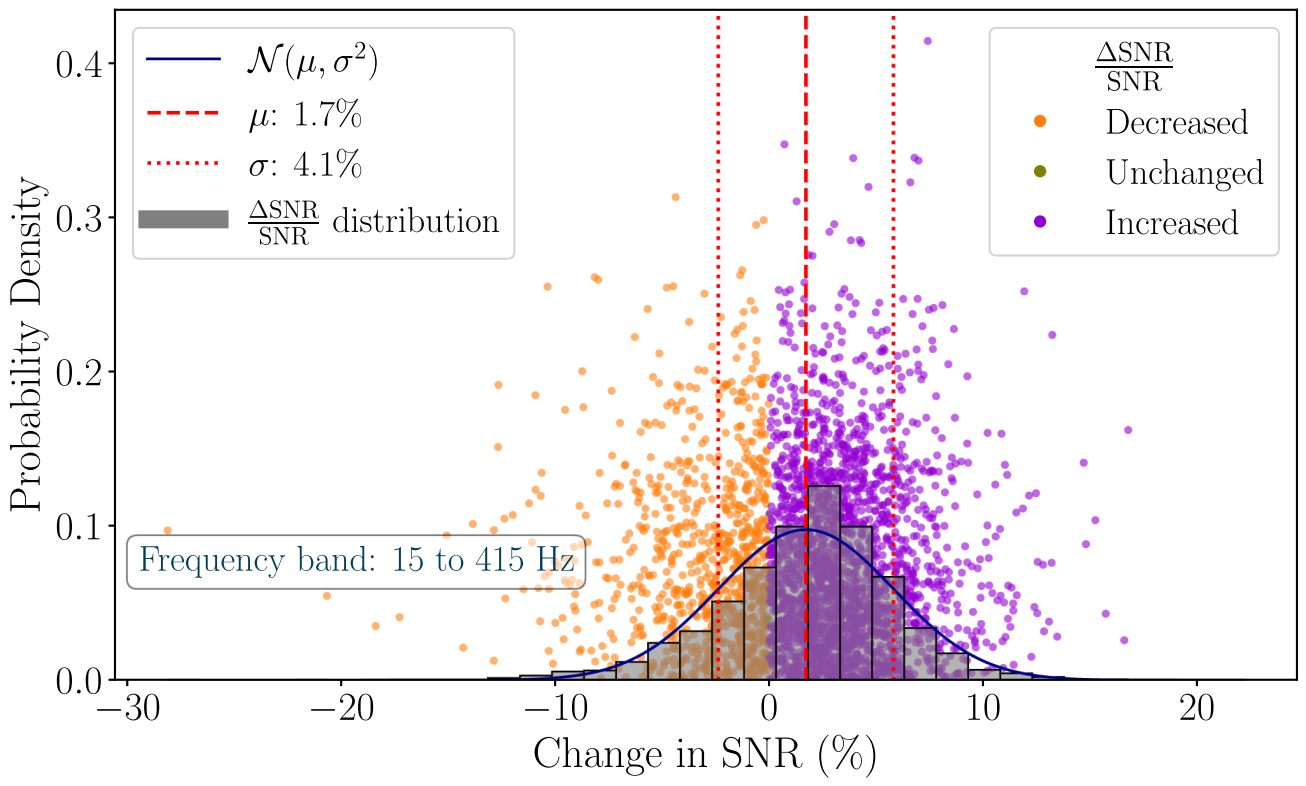

FIG. 8. The figure illustrates the probability density distribution of the relative change in SNR, defined as $\frac{\Delta\text{SNR}}{\text{SNR}}$, for signal injections across the 15–415 Hz frequency band. This compares the SNR values before and after applying `DeepClean`. The histogram bars represent the distribution of the relative change in SNR, with a Gaussian fit overlayed. The key statistics, including the mean ($\mu = 1.7\%$) and standard deviation ($\sigma = 4.1\%$), are highlighted. Individual points depict changes in SNR for each injection and are categorized as decreased, unchanged, or increased.

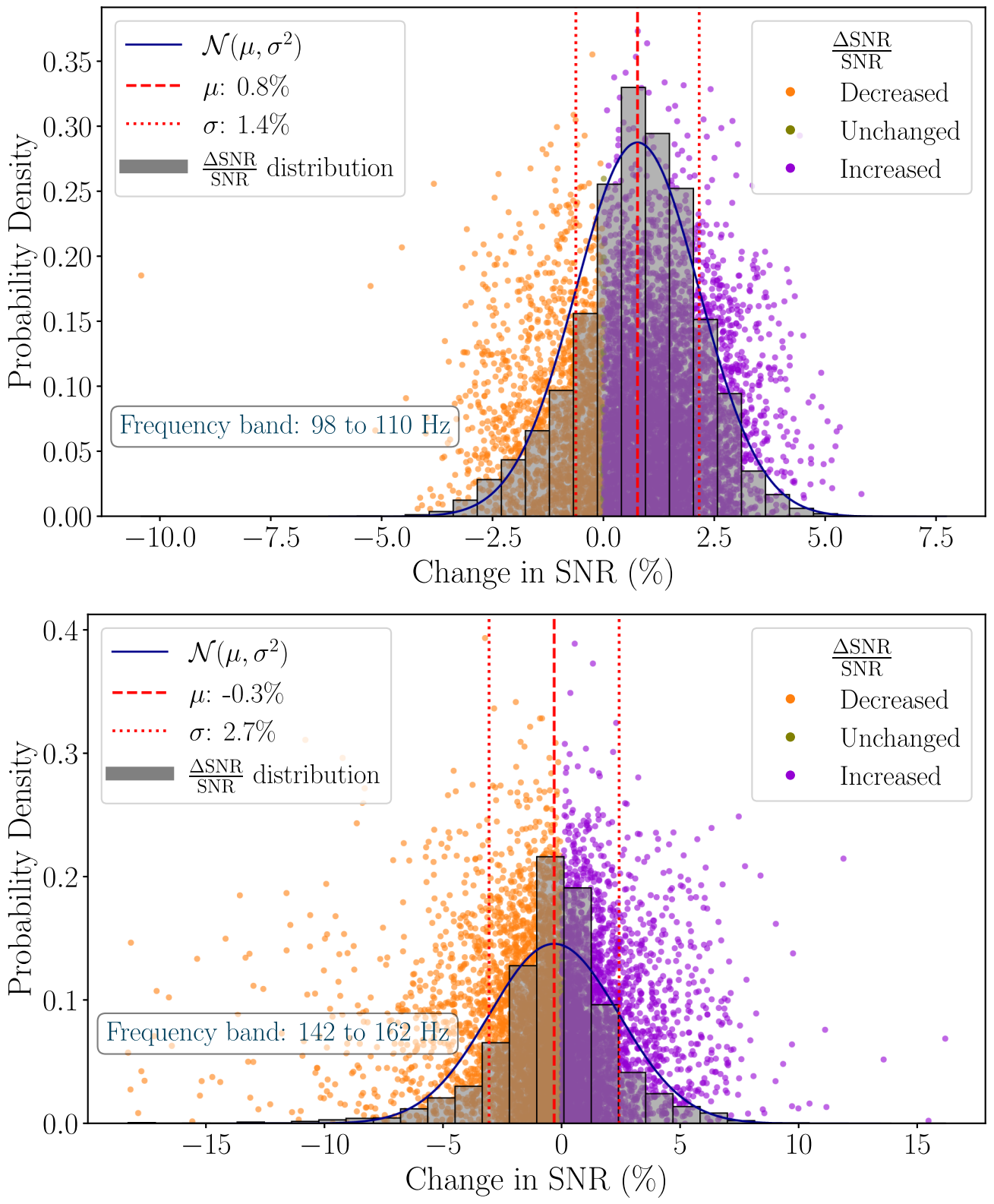

FIG. 9. The top panel illustrates the distribution of the relative change in SNR ($\frac{\Delta\text{SNR}}{\text{SNR}}$) for injections in the 98–110 Hz frequency band before and after applying `DeepClean`. The gray bars represent the probability density of the relative change in SNR, while the solid line shows a Gaussian fit. The mean ($\mu = 0.8\%$) and standard deviation ($\sigma = 1.4$) are indicated with red dashed and dotted lines, respectively. The scatter points represent individual values of the relative change in SNR; orange for decreased, green for unchanged, and violet for increased. The bottom panel presents the same analysis for the 142–162 Hz frequency band. The mean ($\mu = -0.3\%$) and standard deviation $\sigma = 2.7$ are similarly indicated. These figures emphasize the variability and statistical consistency of $\frac{\Delta\text{SNR}}{\text{SNR}}$ after applying `DeepClean` across different frequency bands, with clear categorization of the changes.

While single-band subtractions do not consistently yield a clear SNR gain—with approximately equal numbers of injections gaining and losing SNR—multiband subtraction shows a significant improvement, with over 70% of injections recording an SNR gain.

To analyze these results further, we categorized the SNR fractional differences into three groups based on a threshold: unchanged ($|\frac{\Delta\text{SNR}}{\text{SNR}}| \leq 1\%$), increased ($\frac{\Delta\text{SNR}}{\text{SNR}} > 1\%$), and decreased ($\frac{\Delta\text{SNR}}{\text{SNR}} < -1\%$). For multiband subtraction, 71.3% of GW signals showed an increase, 28.5% showed a decrease, and only 0.2% remained unchanged. For the three single bands—98–110 Hz, 142–162 Hz, and 197–208 Hz—the percentages of injections with SNR gains were 74.8%, 45.3%, and 48.8%, respectively. The remainder in each band either experienced a decrease or remained unchanged (with only 0.2%–3.2% unchanged per band). This test demonstrates that signal recovery by search pipeline remains robust after the application of DeepClean. Even though the gain in SNR is not substantial, we observe a slightly higher number of sources with improved SNR compared to those with reduced SNR.

Table II provides the mean and standard deviations of the SNR fractional difference distributions. The final column includes the $1\sigma$ credible intervals for percentage differences. For multiband subtraction, the $1\sigma$ credible interval indicates that the change in SNR after `DeepClean` relative to the original SNR lies between $-2.4\%$ and 5.9%. The upper bound clearly demonstrates a dominance of SNR gains over losses. For single-band subtractions, as noted earlier, no clear dominance in SNR gain is observed.

TABLE II. Statistical analysis of SNR fractional differences over frequency bands.

| Frequency band | $\mu$ (%) | $\sigma$ (%) | (%) $\frac{\Delta\text{SNR}}{\text{SNR}}$ at $1\sigma$ |
|---|---|---|---|
| | Single training | | |
| 98–110 Hz | 0.8 | 1.4 | −0.62%–2.15% |
| 142–162 Hz | −0.3 | 2.7 | −3.05%–2.42% |
| 197–208 Hz | 0 | 0.4 | −0.35%–0.36% |
| | Multitraining | | |
| 15–415 Hz | 1.7 | 4.1 | −2.36%–5.81% |

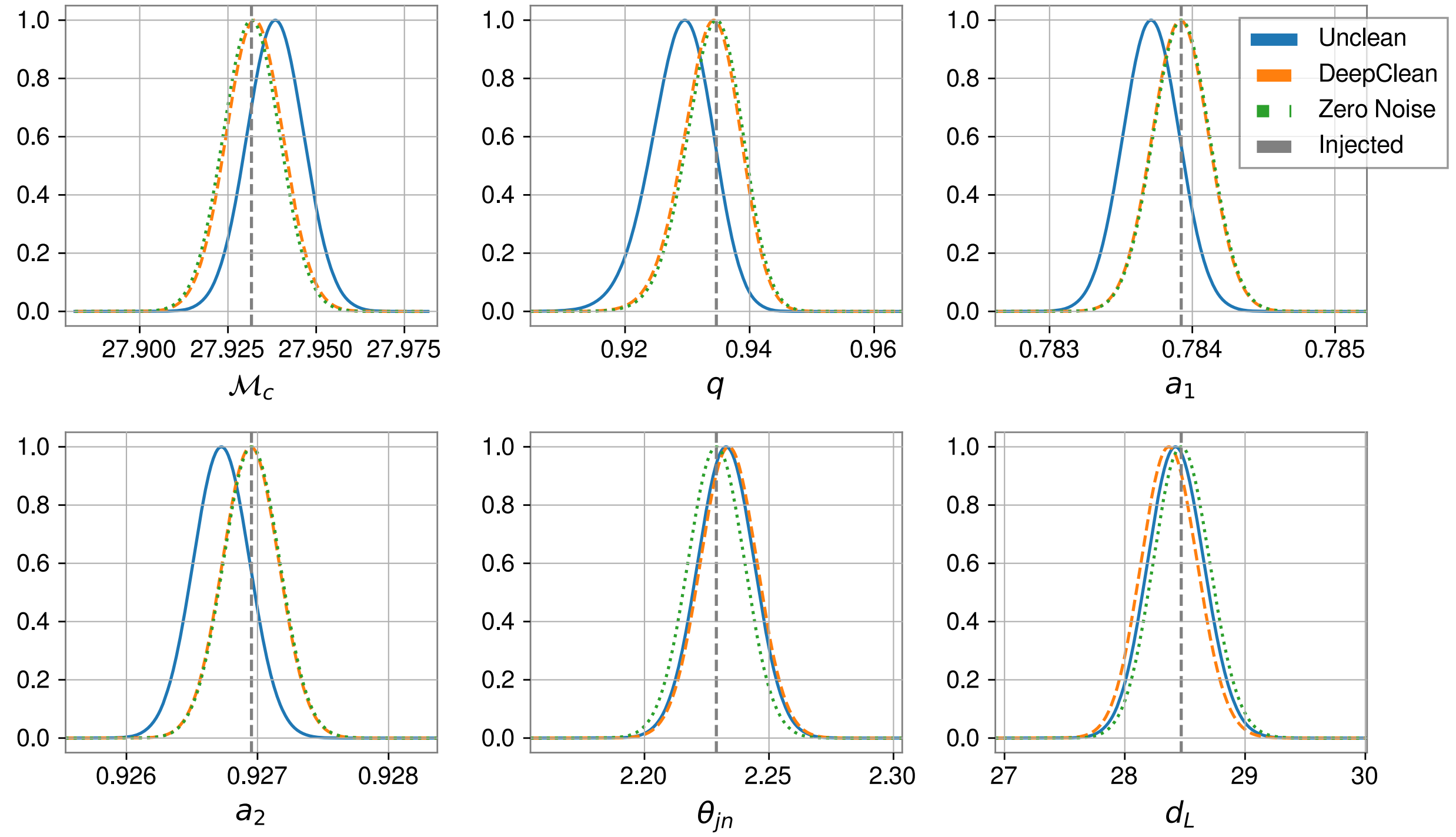

FIG. 10. The posteriors of single-parameter PE (parameter estimation) of one of the 128 injections. The solid (blue) posterior is from the unclean data, the dashed (orange) posterior corresponds to the data after applying `DeepClean`. The injected values are shown in dashed vertical lines (gray). The unclean data yields posteriors peaking away from the true values, which may be attributed to the particular noise realization together with the non-Gaussian artifacts present in the noise. While the `DeepClean` processed data yield posteriors that closely align with the injected values, illustrating reduced noise-induced bias. For sanity of the analysis, we also show the posterior signal with no noise realization (referred to as *zero-noise*) which peaks at the injected values as expected.

### C. Parameter estimation: Testing DeepClean's impact on GW signal structure

To evaluate whether `DeepClean` affects the recovery of astrophysical signals, we perform a set of parameter estimation (PE) studies using injected BBH signals. We adopt two complementary strategies: a *one-at-a-time* approach, where individual parameters are varied while others are fixed at their true injected values, and a full-dimensional analysis, where all 15 BBH parameters are simultaneously estimated.

Each method has distinct advantages. The one-at-a-time approach allows us to isolate the behavior of specific parameters and is particularly useful for detecting subtle biases that might otherwise be absorbed by degeneracies in the full parameter space. For example, if DeepClean introduces additional phase shifts, the effect on the coalescence phase may be redistributed among correlated parameters like time or sky location in a full-dimensional fit. Fixing other parameters ensures that such distortions are directly attributed to the parameter of interest. This method is also computationally efficient and ideal for detecting structural distortions.

In contrast, simultaneous estimation across all 15 parameters yields more realistic posterior uncertainties and provides a holistic view of how DeepClean affects parameter recovery in practice. However, it may obscure small systematic effects due to parameter correlations. Additionally, fixing all but one parameter in the one-at-a-time method may overly constrain the waveform space, preventing the model from exploring alternative morphologies that could better fit a subtly distorted signal. Since neither method is universally more sensitive—each depending on the nature of potential distortions—we perform both to obtain a comprehensive validation.

In the following, we present the analysis and results from both approaches.

#### *1. One-at-a-time parameter estimation*

To assess whether `DeepClean` preserves the structure of astrophysical signals, we perform a validation study using the *one-at-a-time* parameter estimation approach. We injected 128 BBH coalescence signals with ISCO frequencies ranging from 15 to 415 Hz into a 4096 s segment of data, starting at GPS time 1265127585. For each injection, we estimate a single parameter of interest while fixing all others to their true (injected) values, allowing us to isolate DeepClean's impact on that parameter.

In Fig. 10, we present example posterior distributions for the chirp mass $\mathcal{M}c$, mass ratio $q$, component spin magnitudes $a_1$ and $a_2$, inclination angle $\theta_{jn}$, and luminosity distance $d_L$ for one representative injection (injection 24). The blue solid and orange dashed curves represent the posterior distributions obtained from the uncleaned and `DeepClean` strains, respectively, while vertical dashed lines indicate the injected (true) parameter values. We observe that the posterior peaks derived from the uncleaned strain exhibit slight offsets from the injected values due to the random nature of noise. In contrast, posteriors from the `DeepClean` strain align more accurately with both the injected and zero-noise values, demonstrating reduced bias. For comparison, the green dotted curves represent posteriors obtained from the same injection but without any noise realization.

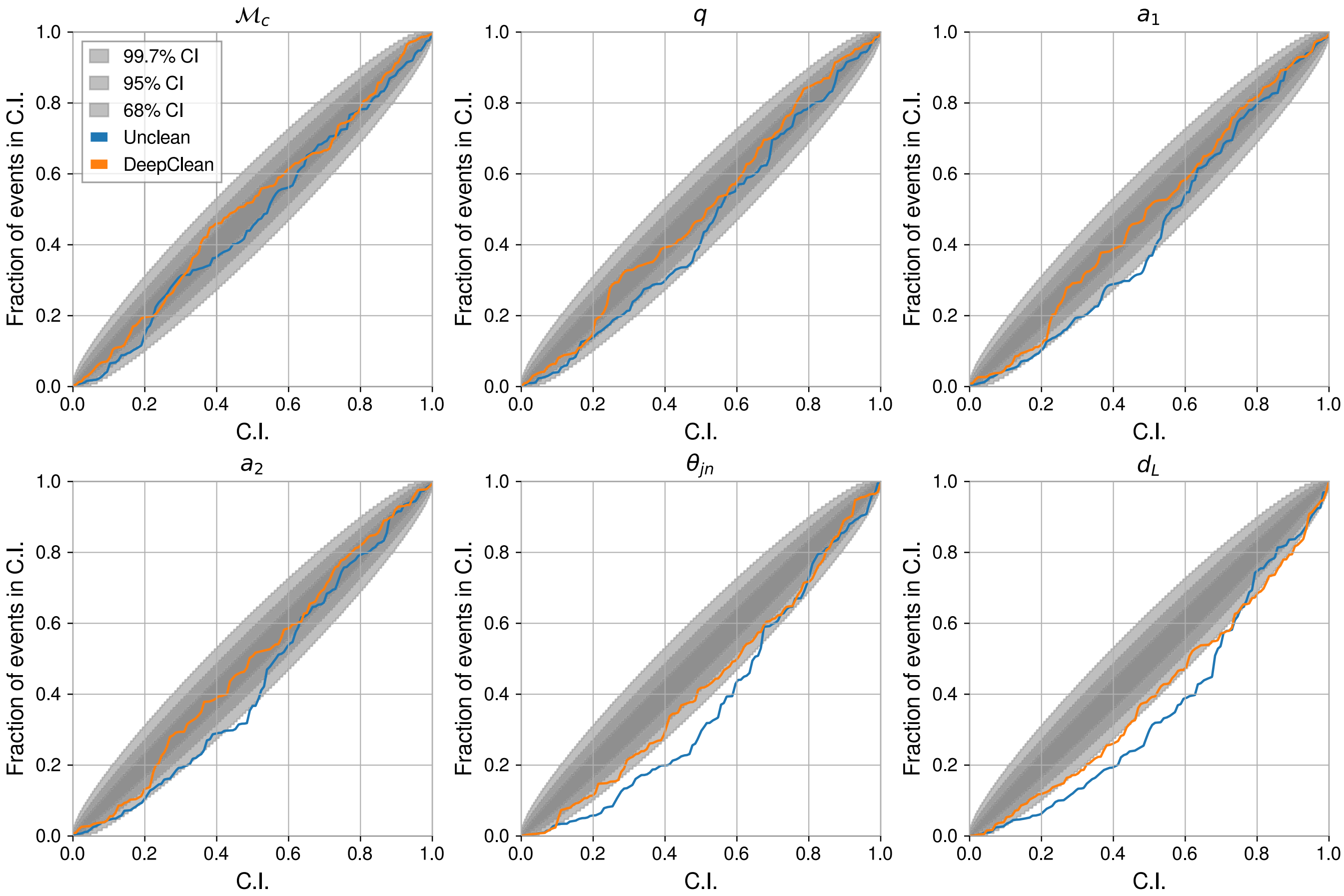

FIG. 11. The PP plots of the six parameters with unclean and `DeepClean` strain showing that `DeepClean` does not affect the recovery of the underlying physical parameters. The shaded regions represent expected statistical fluctuations (68%, 95%, and 99.7% confidence intervals). Combined Kolmogorov-Smirnov p-values are $\simeq 3.47 \times 10^{-14}$ (*Unclean*) and $\simeq 0.0212$ (`DeepClean`). The largest precleaning deviations are for $\theta_{jn}$ ($5.45 \times 10^{-7}$) and $d_L$ ($1.46 \times 10^{-6}$); a milder deviation is seen for the mass ratio $q$ (0.0275). After `DeepClean`, these improve to $\simeq 0.0371$, 0.00585, and 0.679, respectively, while the chirp mass $M_c$ is near-uniform in both cases ($0.361 \rightarrow 0.54$).

Figure 11 presents probability-probability (PP) plots constructed from the full set of 128 BBH injections. These plots illustrate the fraction of injections recovered ($y$-axis) within a specified credible interval ($x$-axis). The results from `DeepClean` closely follow the diagonal reference line and lie within the expected confidence intervals, confirming well-calibrated and unbiased parameter recovery. This indicates that `DeepClean` successfully reduces noise without introducing systematic artifacts into the strain.

However, the PP plots from the uncleaned data show clear deviations, notably falling below the diagonal for parameters such as $\theta_{jn}$, and $d_L$. These deviations indicate biases caused by non-Gaussian noise. These findings underscore that the application of `DeepClean` with a multitraining approach not only efficiently reduces residual noise over a wide frequency range but also preserves GW signal integrity and decreases uncertainties in parameter measurements. These frequency-sensitive parameters also show slight reductions in posterior uncertainties, as illustrated by the narrower posterior distributions in Fig. 10.

### *2. Simultaneous estimation of all 15 parameters*

In this section, we present a parameter estimation (PE) study where all 15 independent parameters are estimated simultaneously.

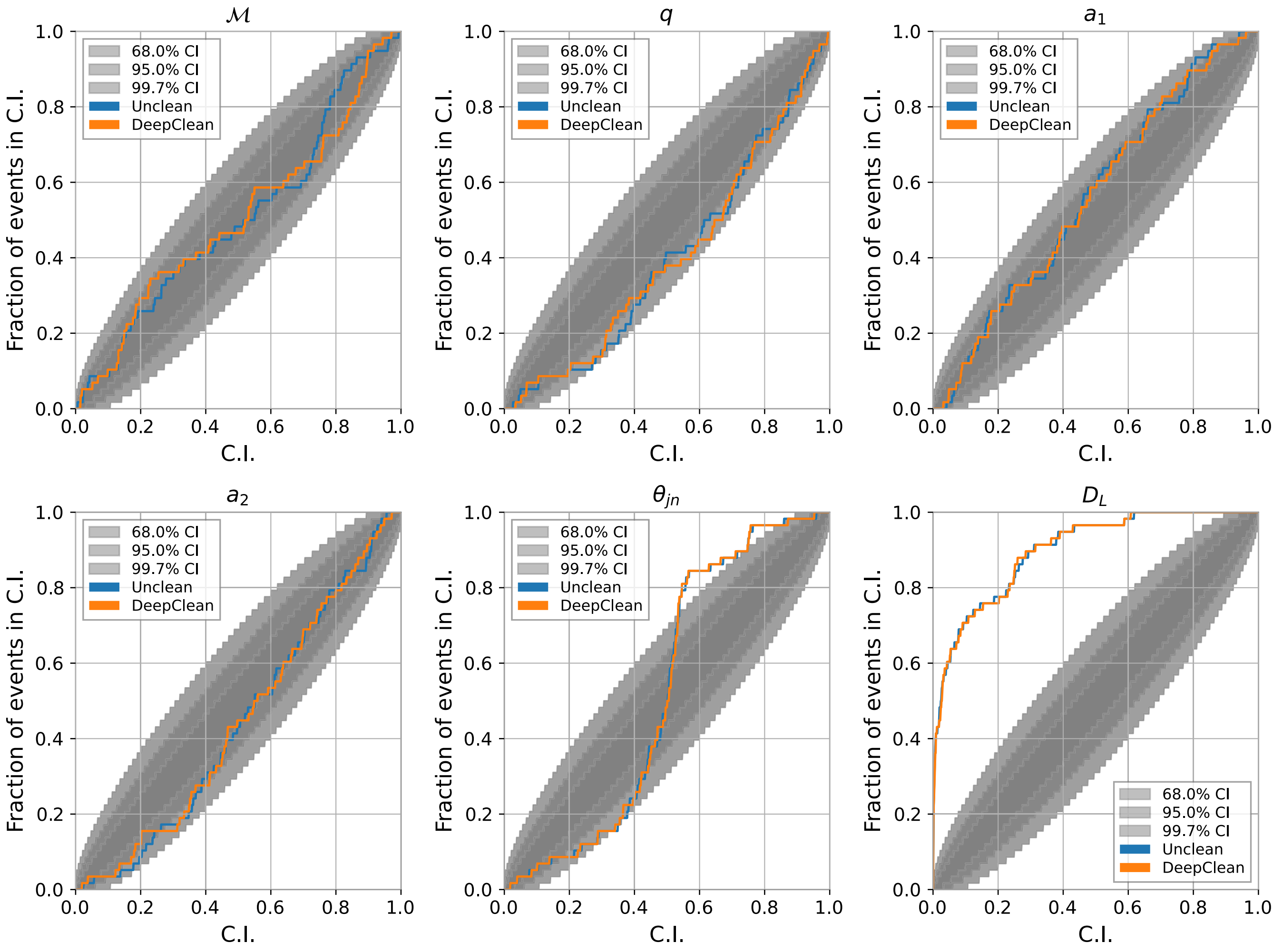

FIG. 12. PP plots for six parameters, comparing *uncleaned* data and data processed with *DeepClean*, using full 15-dimensional parameter estimation. The shaded regions represent expected statistical fluctuations (68%, 95%, and 99.7% confidence intervals). The results indicate that DeepClean preserves the integrity of the GW signal for all parameters except the luminosity distance ($d_L$), which shows a systematic deviation likely due to inherent parameter degeneracies. The combined Kolmogorov-Smirnov p-values are $\simeq 2 \times 10^{-4}$ for the *Unclean* data and $\simeq 1 \times 10^{-4}$ after *DeepClean*.

We consider the same injections as in the *one-at-a-time* analysis but analyze only 60 signals due to computational reasons. Note that this is a single-detector analysis, and hence it is difficult to constrain extrinsic parameters such as time and phase of arrival, luminosity distance, and the angles describing the source's location and orientation. In Fig. 12, we show the PP plots for chirp mass, mass ratio, the two spin magnitudes, inclination angle, and luminosity distance. As expected, the PP plots for inclination angle and luminosity distance show poor recovery of the true values, mostly reflecting the priors. However, for masses and spins, the PP plots indicate good and comparable recovery both before and after applying DeepClean. There is no evidence of any bias introduced by the application of DeepClean.

While the PP plot can be used as a measure of credibility of the injection recovery, it alone does not show whether the noise subtraction has improved the error bars. For that matter, in Fig. 13, we compare the width of the 90% credible intervals for the chirp mass (left) and the mass ratio (right). The histograms show the ratio of the interval width after and before applying DeepClean, i.e., V1:DC/V1:ORG, implying that a value below 1 should indicate a reduction in uncertainty.

Even in cases where DeepClean effectively performs no subtraction, i.e., when the predicted noise is negligible, there is still an inherent randomness in the predicted noise realization. This is due to both the stochastic nature of the targeted noise coupling and the finite-duration nature of the regression analysis. This adds an additional layer of uncertainty to the cleaned data, and can reflect in the analysis data where the BBH signals are injected, as well as the PSD estimated from neighboring segments. These fluctuations can, in turn, cause small variations in the posterior widths. In an idealized scenario with no real subtraction, one might expect the error bar ratios (after vs before DeepClean) to form a delta function at 1. However, in practice, due to the reasons outlined above, the distribution instead peaks around 1 but with spread on either side.

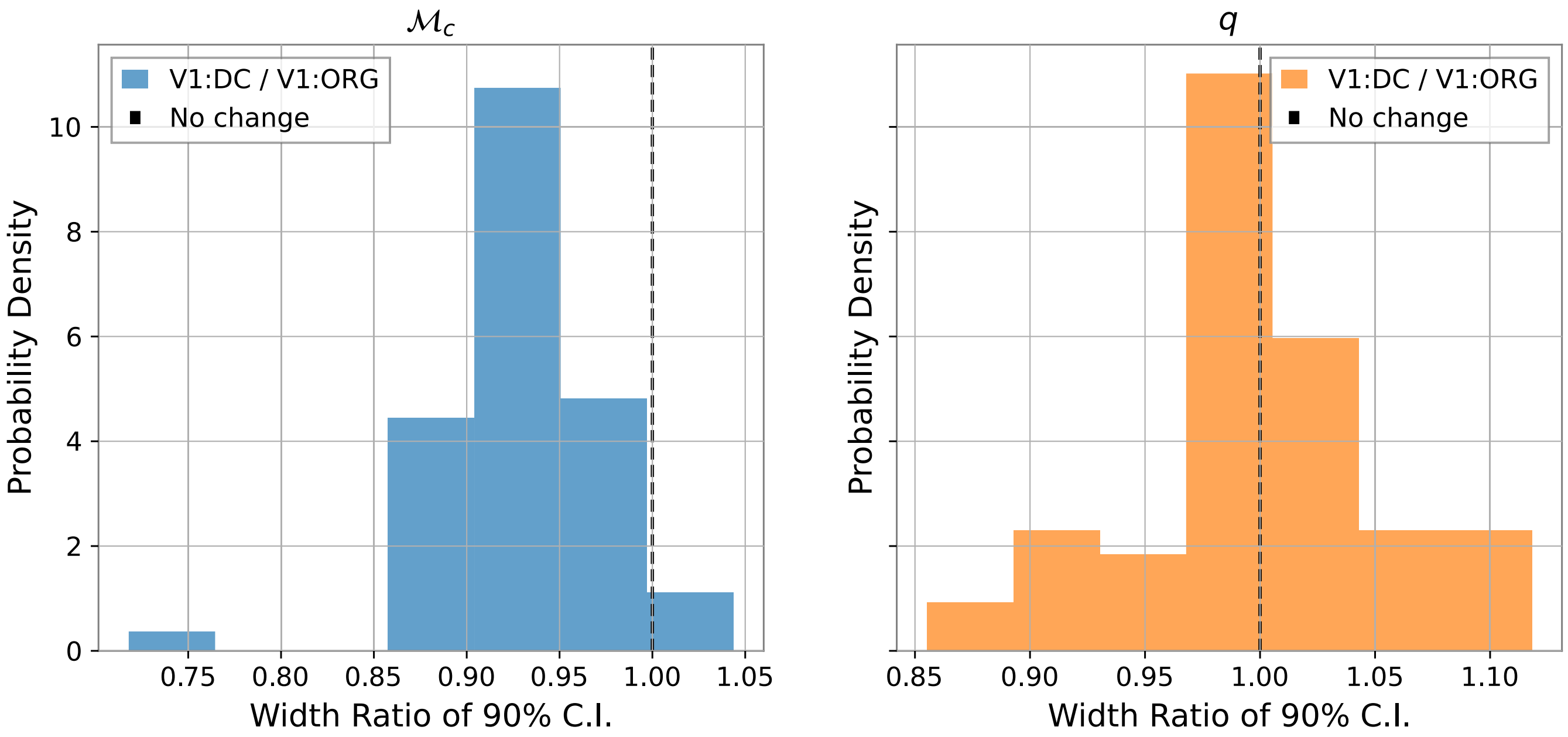

FIG. 13. Histogram of error bar ratios (DeepClean/original) for chirp mass (left) and mass ratio (right). Chirp mass shows a clear reduction in uncertainty, while mass ratio reflects neutral behavior with symmetric fluctuations around unity.

This is precisely what we observe for the mass ratio in the right panel of Fig. 13; the histogram of error bar ratios is centered near 1 with roughly symmetric fluctuations. This behavior is consistent with a scenario where DeepClean does not yield significant improvement for this parameter and should not be interpreted as evidence of any adverse effect from the subtraction. In contrast, for the chirp mass—the most sensitive parameter in an inspiraling compact binary merger—we observe a clear shift toward ratios less than 1, indicating a reduction in uncertainty. In summary, both parameters exhibit a spread in their error bar ratio distributions due to the randomness in the noise predictions, with chirp mass showing an overall improvement and mass ratio reflecting neutral behavior.

## V. SUMMARY AND CONCLUSIONS

In this paper, we applied the `DeepClean` framework to the Virgo detector data and demonstrated the successful subtraction of nonlinear noise couplings across a wide frequency range. Between 15 and 415 Hz, we divided the strain data into 13 nonoverlapping frequency bands and performed noise regression sequentially across them. Through this study, we demonstrate the effectiveness of nonlinear noise regression in Virgo using `DeepClean`, resulting in improved detector sensitivity and enhanced detection prospects for GW signals. We have noted up to a 1.3 Mpc improvement in the BNS inspiral range (∼2.5% gain).

Importantly, our validation tests show that the astrophysical signals are not adversely affected by the regression process, as confirmed through parameter estimation consistency checks. Additionally, we observed an increase in matched-filter SNR for multiple injected signals, which revealed an average increase of 1.7% in the recovered SNR for injected BBH. These results suggest that such regression methods can contribute meaningfully to future observing campaigns.

While `DeepClean` has been previously demonstrated with the two LIGO detectors, showcasing its capability for nonlinear noise regression in Virgo—a detector with distinct design features—carries importance as a proof for their generalizability beyond their original deployment. Specifically, it validates the broader applicability of such tools and serves as a test for several Virgo-specific aspects that may challenge noise-regression efforts, such as channel availability, coupling stability, and noise behavior.

One of the longstanding challenges in noise regression has been the presence of a large number of witness channels, many of which exhibit non-negligible coherence with the strain channel. Traditional filtering techniques often struggle to process more than a few tens of such channels due to analytical and computational complexity. In contrast, deep-learning-based approaches such as `DeepClean` offer a scalable alternative. In this work, even within our band-split strategy, we have one case where 122 witness channels were used simultaneously in a single-pass training (see Fig. 7), and this configuration outperformed the multilayer strategy that grouped channels by category within that particular band. This not only confirms `DeepClean`'s capacity to handle large numbers of channels within memory constraints, but also highlights the presence of meaningful cross-category couplings that are best captured when training is performed jointly. While there are practical memory limits, our analysis demonstrates a level of channel multiplicity far beyond what traditional methods can handle.

While our initial selection of witness channels was based on Brute Force Coherence—primarily identifying linear noise couplings—`DeepClean` inherently enables the removal of nonlinear and nonstationary noise couplings that may exist between those channels, as we have shown in Sec. III E. However, identifying a complete set of witness channels capable of capturing all such nonlinear couplings goes beyond what Brute Force Coherence can identify. This would require a systematic exploration and validation of a broader set of witness channels, extending beyond linear coherence. Addressing this challenge is an open problem and is beyond the scope of the present work.

In the future, there are a number of challenges we plan to tackle:

(i) Online analysis: All analysis in this paper was performed offline. However, online processing will be critical in O5 to leverage noise improvements in real time and enhance premerger GW detection. To support this, we plan to transition from the sequential bandwise cleaning strategy used in this work—where the output of one band feeds into the next—to a parallel bandwise architecture. This will allow training and cleaning across multiple frequency bands simultaneously, significantly improving computational efficiency and reducing latency. The core algorithmic components for this parallelization have already been developed and tested on LIGO data, and can be readily adapted for Virgo. We also iterate that this parallelization occurs across frequency bands (not across witness channels), and as long as these bands remain nonoverlapping, there is no conflict even if some channels contribute to multiple bands.

(ii) High computational demands: Capturing the 150 Hz peak and its sidebands using ASC witness sensors required considerable computational resources, often requiring tradeoffs between data length and resolution. We plan to explore lighter versions of the architecture for simpler forms of coupling, as well as alternative training strategies to reduce memory usage. For instance, all relevant noise couplings in this study are below 500 Hz, suggesting that a sampling rate of 1 kHz or lower may be sufficient. Nevertheless, our current pipeline operates at 4096 Hz to match astrophysical data needs. With careful tuning, lower sampling rates in the training phase may help optimize performance without compromising scientific goals.

(iii) Data access: To mitigate data read latency, we explored the VIRGOTOOL Python package.[6] However, compatibility issues with our cluster limited its use. We recommend integrating VIRGOTOOL's data handling capabilities into the more widely adopted GWPY library to streamline data access and improve computational efficiency in future studies.

(iv) Configuration: Moving from LIGO to Virgo required only configuration changes, mainly in the choice of witness channels, frequency-band setup, and training parameters, while the `DeepClean` algorithm itself remained unchanged. This demonstrates its transferability.

## ACKNOWLEDGMENTS

R. W. K. were supported by the Axe "ondes gravitationnelles/multimessager" de l'Observatoire de la Côte d'Azur, Boulevard de l'Observatoire, F-06304 Nice. R. W. K. acknowledge ANR (French National Research Agency) for its support of the project "Multimessenger observations of the Transient Sky (MOTS)" No. ANR-22-CE31-0012. M. S. and M. W. C. acknowledge the support from the National Science Foundation with Grants No. PHY-2308862 and No. PHY-2117997. M. S. also acknowledges the support from Weinberg Institute for Theoretical Physics at the University of Texas at Austin. The authors thank Franco Carbognani for making the O3b Virgo data available on the CIT cluster for our work. We are also very grateful to Alba Romero Rodríguez for her review on behalf of Virgo DRS and LIGO P&P, which has been useful in the improving of this paper. We would also like to thank Nathalie Besson for her valuable feedback, which enhanced the clarity and precision of the manuscript. The authors acknowledge the International Gravitational-wave Network (IGWN) Computing Grid (CIT, LHO, LLO) for providing resources to realize our simulations. We acknowledge the Virgo DetChar and Virgo cluster at Cascina for providing resources that contributed to the research results reported within this paper. This research used resources of the National Energy Research Scientific Computing Center (NERSC), a U.S. Department of Energy Office of Science User Facility operated under Contract No. DE-AC02-05CH11231 under Project "Toward a complete catalog of variable sources to support efficient searches for compact binary mergers and their products." This work used the Advanced Cyberinfrastructure Coordination Ecosystem; Services and Support (ACCESS). This work used the Extreme Science and Engineering Discovery Environment (XSEDE) COMET at SDSU through Allocations No. AST200016 and No. AST200029.

[6] https://anaconda.org/conda-forge/pythonvirgotools

This material is based upon work supported by NSF's LIGO Laboratory, which is a major facility fully funded by the National Science Foundation.

## DATA AVAILABILITY

The data that support the findings of this article are openly available [43].

## APPENDIX

### 1. Popular subtraction techniques

Effective noise reduction in GW detectors is pivotal for enhancing the signal detectability as well as the astrophysical outcome. A variety of algorithms, including traditional approaches and those based on machine learning, are now present in the literature for this purpose:

*Wiener filtering*. This is a fundamental filtering technique that is effective when the noise is *stationary* and noise coupling is linear. It aims at minimizing the mean square error (MSE) between estimated and true signals [12,21]. Its robustness is evident in GW event analysis by the IGWN [22]. However, its performance becomes sub-optimal for *nonstationary* and *nonlinear* noise driven by complex environmental and instrumental fluctuations. This highlights the need for adaptive noise reduction strategies in GW astronomy.

*NonSENS*. Nonstationary Estimation of Noise Subtraction (NonSENS) employs an Infinite Impulse Response (IIR) filter to remove both stationary and nonstationary noise in GW detectors [14]. This method effectively addresses noise arising from slow interferometer movements, such as angular fluctuations, which pose challenges to traditional techniques. By utilizing witness signals to monitor noise and its modulation over time, NonSENS achieves stable, parametric subtraction of *nonstationary* noise. It leverages data science techniques, such as gradient-based optimization [31], to estimate the coupling coefficients, thereby overcoming the complexities that hinder standard noise reduction methods. Additionally, the analytical filter, with its limited number of coefficients, avoids the computational challenges and interpretability issues often associated with deep neural networks.

*Adaptive Feed-Forward*. During the O3 run, Advanced Virgo implemented an adaptive linear feedforward technique to suppress 50 Hz noise originating from the electrical mains [20]. A phase signal from the Uninterruptible Power Supply (UPS) in the Central Building serves as a witness channel for this disturbance. The signal is filtered through a narrow-band resonant filter centered at 50 Hz and adjusted in amplitude (gain) and phase to match the coupling of the noise into the differential arm length (DARM) channel. The resulting correction is then injected into DARM, effectively cancelling the targeted line. This control is adaptive: the gain and phase are continuously updated in real time based on demodulation of DARM with respect to the UPS signal, allowing the system to track variations in the noise coupling path and maintain robust suppression of the 50 Hz line.